\documentclass[prb,twocolumn,showpacs,floatfix,superscriptaddress,epsf,psfig]{revtex4}
\usepackage{epsf}
\usepackage{color}
\usepackage{graphicx}
\usepackage{amsmath, amssymb}
\usepackage{lscape}
\usepackage{rotating}
\begin{document}
\title{Massless Dirac-Fermions in Stable Two-Dimensional Carbon-Arsenic Monolayer} 
\author{C. Kamal}
\affiliation {Theory and Simulations Laboratory, HRDS, Raja Ramanna Centre for Advanced Technology, Indore - 452013, India}
\affiliation {Homi Bhabha National Institute, Training School Complex, Anushakti Nagar, Mumbai-400094, India}

\begin{abstract}
We predict from density functional theory based electronic structure calculations that a monolayer made up of Carbon and Arsenic atoms, with a chemical composition (CAs$_3$) forms an energetically and dynamically stable system.  The optimized geometry of the monolayer is slightly different from the buckled geometric configuration observed for silicene and germanene. The results of electronic structure calculations predict that it is a semi-metal. Interestingly, the electronic band structure of this material possesses a linear dispersion  and a Dirac cone at the Fermi level around the high symmetric $K$ point in the reciprocal lattice. Thus, at low energy excitation (up to 105 meV), the charge carriers in this system behave as massless Dirac-Fermions. Detailed analysis of  partial density of state suggests that the 2p$_z$ orbital of C atoms plays vital role in determining the nature of the states, which has a linear dispersion and hence the Dirac cone, around the Fermi level.  Thus, the electronic properties of CAs$_3$ monolayer are similar to those of graphene and other group IV based monolayers like, silicene and germanene.  In addition, we have also investigated the influence of mechanical strain on the properties of CAs$_3$ monolayer. The buckled configuration becomes the planar configuration for a tensile strain beyond 18\%.  Our results indicate that the monolayer possesses linear dispersion in the electronic band structure for a wide range of mechanical strain from -12 to 20\%, though the position of Dirac point may not lie exactly at the Fermi level. The linear dispersion disappears  for a compressive strain beyond -12\%  and it is due to the drastic changes in the geometrical environment around C atom which in turn modify the hybridization between its orbitals.
Finally, we wish to point out  that CAs$_3$ monolayer belongs to the class of \textit{Dirac materials} where the behaviour of particles, at low energy excitations, are characterized by the Dirac-like Hamiltonian rather than the Schrodinger Hamiltonian.

\end{abstract}

\maketitle

\section{Introduction}
Research on two-dimensional (2D) materials has increased tremendously after the discovery of  graphene - an atom-thin honeycomb monolayer of  carbon. Graphene is considered as a \textit{wonder material} because it possesses many novel  properties such as the charge carriers in this material behave as Dirac-Fermion, Klein tunneling, anomalous half-integer quantum hall effect, finite DC conductivity,  etc\cite{graphene,graphene1}.  Researchers around the world have been searching for new graphene-like 2D materials made up of different  elements, in particular from p-block of the periodic table. There has been significant progress in this direction. In the last one decade, several new 2D monolayers have been theoretically predicted to be stable\cite{boro1, boro2,boro3, alum1, sili1, sili2,sili3, sili-ck1,sili-ck2,sili-ck3, group4_1,group4_2, germ1,sn1, plum1, nitr1,phos1,phos2,phos3, arse1, anti1,bism1} and later quite a few of them have been successfully grown  experimentally\cite{Eboro1,Eboro2,Esi1,Esi2,Esi3, Esi4,Ege1,Ege2,Ege3,Esn1,Esn2,Esn3,Eph1, Eph2,Eas1,Egroup5_1,Egroup5_2}.  In analogy with graphene, the 2D monolayers based on p-block elements (given in parenthesis) are named as  borophene (B), aluminene (Al),  silicene (Si), germanene (Ge), stanene (Sn), plumbene (Pb),  phosphorene (P), arsenene (As), antimonene (Sb), bismuthene (Bi), etc. From the numerous studies available in the literature, many interesting properties of the above mentioned 2D materials have been revealed. Some of the important outcomes from these studies\cite{boro1, boro2,boro3, alum1, sili1, sili2,sili3, sili-ck1,sili-ck2,sili-ck3, group4_1,group4_2, germ1,sn1, plum1, nitr1,phos1,phos2,phos3, arse1, anti1,bism1,Eboro1,Eboro2,Esi1,Esi2,Esi3, Esi4,Ege1,Ege2,Ege3,Esn1,Esn2,Esn3,Eph1, Eph2,Eas1,Egroup5_1,Egroup5_2} which are relevant to the present paper are given in the following.

Similar to graphene,  the group IV based monolayers such as silicene, germanene, stanene are found to be semi-metal and all of them possess a Dirac cone at the Fermi level (E$_F$) around the high symmetric $K$ point in the reciprocal lattice. Hence the charge carriers in these monolayers behave as massless Dirac-Fermion at their low-energy excitation. It is important to note that these group IV based monolayers stabilize in a buckled geometric configuration which is different from the planar geometry of graphene. However, the buckling in the geometry does not alter the presence of Dirac cone at the E$_F$. It has been reported in the literature that the buckling will play an important role in inducing a band gap when the monolayers are subjected to external transverse electric field\cite{sili-gap1,sili-gap2,sili-gap3,sili-gap4,sili-ck2}.  It is well known that many properties of materials are determined by the nature of states around the Fermi level. In case of the group IV based 2D materials, the properties of materials at low-energy excitation are characterized by the Dirac-Fermion (the linear dispersion around the E$_F$) unlike the usual Schrodinger-Fermion (parabolic-like dispersion). Due to this reason, the group IV based 2D monolayers, including graphene, belong to a class of material called \textit{Dirac materials}\cite{dirac} in which the low-energy excited carriers follow the Dirac-like Hamiltonian.   Until now, many different materials have been shown to possess particles which behave as Dirac-Fermion and consequently many of their  properties at the low-energy excitations are universal\cite{dirac}.

On the other hand, the studies on the properties of  group III based 2D materials such as aluminene and borophene show that these are metals whereas the group V monolayers (phosphorene, arsenene, antimonene, bismuthene) behave as semiconductors.  The electronic band structure calculations  of free standing borophene show that it is a highly anisotropic metal due to crossing of bands along $\Gamma-X$ and $S-Y$ directions\cite{Eboro1}. In case of aluminene, the electronic band structure resemble very close to that of graphene, but the Dirac cone lies well  above the E$_F$ (by the amount 1.618 eV)\cite{alum1}. This is due to fact that Al atom is trivalent as compared to tetravalency of C atom and hence the $p_z$ orbital is not completely filled.  It is clearly observed that the monolayer has a finite DOS at the Fermi level and hence shows interesting Fermi curves\cite{alum1}. In contrast, phosphorene is a direct band semiconductor and it has very high hole mobility\cite{phos1}. In addition, a band gap can be tuned by varying the number of layers\cite{phos1}. Hence, these results are considered to be important from the applications point of view. Few devices based on phosphorene have also been demonstrated\cite{Eph1}. Another group V based monolayer - arsenene is found to be indirect band gap semiconductor, but it goes from indirect-to-direct due to applications of very small amount of mechanical strain\cite{arse1}. Thus, this can also be a potential candidate for many optoelectronic applications. Moreover, semiconducting behaviour is also observed in case of antimonene and bismuthene monoalyers\cite{anti1,bism1}.  Apart from the elemental monolayers, investigations on the properties of binary monolayers have also been reported in the literature. Sahin et. al. have carried out detailed investigation on several properties of group III-V  based binary monolayers which are all found to be semiconducting in nature\cite{group3-5}.  Similarly, computational study  on the electronic properties of group IV-VI based monolayer reveals that they are also semiconductors\cite{group-4-6}. In contrast to metallic character of GeP$_3$\ and SnP$_3$ bulk systems, the monolayer counter parts of these two group IV-V based systems are found to be semiconductors\cite{gep3,snp3-1,snp3-2}

We note here that all the binary monolayers made up of p-block elements as well as the elemental monolayers of group III and V elements reported in the literature  are either semiconductor or metal.  Thus, they do not exhibit properties similar to those of graphene and other group IV based monolayers which are all semi-metal and possess Dirac cone at the E$_F$. 
Since the Dirac materials show many exciting properties, it is important to explore the possibility of discovering different types of new graphene-like 2D monolayers  which may possess massless Dirac-Fermions. This will be expected to widen the scope of the research in 2D materials because many of these materials are considered as potential candidates for several technological applications, in particular in the field of flexible electronics due to their excellent mechanical and  electronic properties\cite{flex-2d}. Thus, searching for new 2D Dirac materials is very important from both the fundamental as well as application points of view.  

With the  above mentioned motivations, we explore the possibility of  finding a stable binary monolayer made up of group IV and V elements, namely C and As atoms respectively.  For this purpose, we study the stability and various physical properties such as geometric, electronic and vibrational properties of  Carbon-Arsenic  (CAs$_3$) monolayer by employing density functional theory (DFT) based electronic structure calculations. 
Results of our phonon and cohesive energy calculations show that CAs$_3$ monolayer in buckled configuration is both energetically and dynamically stable. Like graphene, it is a semi-metal and possesses a linear dispersion and a  Dirac cone at the E$_F$ around the high symmetric $K$ point in the reciprocal lattice. Thus, the charge carriers in this material behave as massless  Dirac-Fermion. The energy range through which the Dirac cone exists is from  nearly -600 meV to 250 meV. However, due to the presence of  a parabolic-like dispersion curve  at $\Gamma$ point around 105 meV above the E$_F$, the characteristic of Dirac-Fermion like behaviour of electron-like particle is limited to excitation up to 105 meV.  For excitation beyond this energy,  the contributions from Schrodinger-like particles will also be included.  In addition, we have also studied the influence of mechanical strain on the geometric and electronic properties of the CAs$_3$ monolayer. Our results suggest that there is a transition from the buckled to a planar configuration when an applied tensile strain goes beyond 18\%.  For a wide range of mechanical strains (-12 to 20 \%), the monolayer possesses the linear dispersion, though the position of the Dirac point does not lie exactly at the E$_F$.  Detailed analysis indicates that the drastic changes in the geometrical environment around C atom for a compressive strain beyond -12\% cause the linear dispersion to disappear.  

The paper has been organized in the following manner. Next section contains the details of computational approach employed in the present study. It is then followed by results and discussion in Section III.  Finally, we conclude our results in the last section. 

\section{Computational Details}

\quad
We employ Vienna ab-initio simulation package (VASP)\cite{vasp} within the framework of the projector augmented wave (PAW)\cite{paw} method to perform density functional theory (DFT)\cite{dft} based electronic structure calculations.  For exchange-correlation (XC) functional, we choose  generalized gradient approximation (GGA) given by Perdew-Burke-Ernzerhof (PBE)\cite{pbe}.  The plane waves are expanded with energy cut-off of 400 eV. We use $\Gamma$ centered mesh of  31$\times$31$\times$1 for k-point sampling for  Brillouin zone integrations.   The convergence criteria for energy in SCF cycle is chosen to be 10$^{-6}$ eV.  The geometric structure is optimized by minimizing the forces on individual atoms with the criterion that the total force on each atom is below 10$^{-2}$ eV/\AA{}.  A vacuum of about 18 \AA {} is applied in the direction perpendicular to the plane of 2D sheet to make the interaction between two adjacent unit cells in the periodic arrangement is negligible. Drawing of geometry and charge density has been done using VESTA software\cite{vesta}.
For phonon calculations, we use small displacement method as implemented in phonopy code. In order to calculate force constant, few atoms have been displaced in a supercell 3$\times$3$\times$1 which contains 72 atoms. 

\section{Results and Discussion}
 \subsection{Geometric Structure and Stability}
\begin{figure}[!t]
\begin{center}
\includegraphics[width=0.5\textwidth]{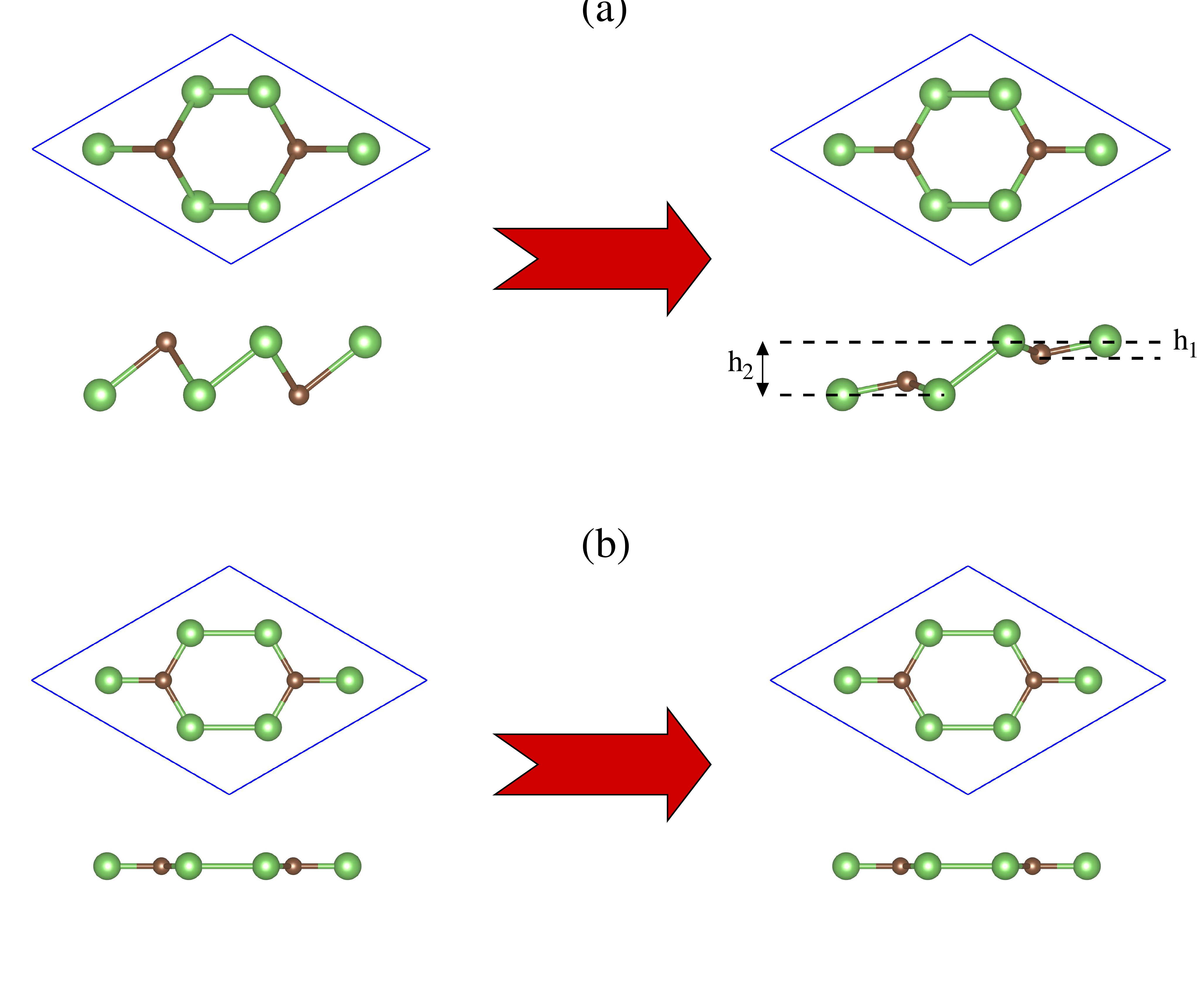}
\end{center}
\caption{ (color online) Top and side views of  initial and optimized geometric structures of CAs$_3$ monolayer in (a) buckled and (b) planar configurations. The vertical distances h$_1$ and h$_2$ represent the buckling lengths in the optimized geometry of buckled  configuration.}
\label{FigStruct}
\end{figure}

Figure 1 shows the initial and optimized geometric structures of CAs$_3$ monolayer in two different configurations namely (a) buckled and (b) planar. These initial geometries are similar to those of silicene and graphene in  2$\times$2$\times$1 supercell respectively. The unit cell of CAs$_3$ monolayer contains 2 carbon and 6 arsenic atoms. Results of the optimized  geometrical parameters (lattice constant, bond lengths and bond angles) and binding energies for these two configurations are summarized in Table I.  We have observed that the planar configuration retains its geometry whereas the buckled structure is converged to a slightly different geometry. The value of lattice constant for buckled geometry is found to be 6.778 $\AA$ which is much shorter than that (7.964 $\AA$) of planar configuration.  Neighboring atoms in the initial geometry of buckled configuration (as well as in silicene-like geometry) lie in two different planes and the vertical distance between the atoms in these two planes is called buckling length.  But, in the optimized geometry of CAs$_3$ monolayer,  C and As atoms have moved to two different sets of planes and thus overall two buckling distances, namely h$_1$ and h$_2$ are required to characterize the buckling in the system: the former is the vertical distance between a C atom and its neighboring As atom and the latter is the vertical distance between two neighboring As atoms.  It is to be noted that when h$_1$ and h$_2$ become zero,  the buckled structure will reduce to planar structure. Our calculated values of  As-C and As-As bond lengths for buckled configuration are 1.920 and 2.565  $\AA$ respectively. However, reduction of As-C bond (1.897 $\AA$) and extension  of As-As bond (2.701 $\AA$) have been observed for the planar configuration. The characteristic bond angle of 120$^\circ$ between the constituent atoms is retained in honeycomb lattice of planar configuration.  In case of buckled configuration,  three different types of bond angles,  namely As-C-As (116.09$^\circ$), As-As-As (86.64$^\circ$) and C-As-As  (105.41$^\circ$) have been observed. We wish to note here that the bond angles centered around As atoms are much smaller as compared to those around C atoms. Moreover, the values of bond angles around the former and latter atoms are close to 109.47$^\circ$ and  120$^\circ$ respectively. This clearly indicates that As and C atoms prefer to be in sp$^3$-like and sp$^2$-like hybridizations.  Buckling distance (h$_1$) between C and As atoms is found to be 0.385 $\AA$. Further, the lower value of bond angle between As atoms  and the higher value of buckling  length (h$_2$ = 1.565 $\AA$) as compared to those observed in buckled arsenene (92.22$^\circ$, 1.388 $\AA$)\cite{arse1} clearly indicate that the effect of buckling is stronger in CAs$_3$ monolayer rather than that in buckled arsenene.

\begin{table}[]
\footnotesize
\begin{center}
\caption{The results for the binding energy, lattice constant, bond length and bond angle of CAs$_3$ monolayer obtained from DFT based calculations with GGA XC functional.}
 \begin{tabular}{lcccc}
\hline
\hline
Symmetry & 	Binding 		&	Lattice 	&	Bond  	& Bond		\\
              &	Energy	      	&	Constant &	Length	& Angle 		\\
	      &	$E_B$ (eV/atom)	&	 $a_0$ (\AA{}) 	&	(\AA{})	&  ($^{\circ}$)	\\
\hline
P-3m1	      &	 -3.584  		&6.778 &	1.920 (As-C)	& 116.09 (As-C-As)	\\
(buckled) &	   	&	&	2.565 (As-As)& 86.64 (As-As-As)	\\
	      &	    	&		&				& 105.41 (C-As-As)	\\
P6/mmm &-3.258 & 7.964& 1.897 (As-C)&  120 (As-As-As)\\
(planar)		&		&		&2.701 (As-As)& 120 (C-As-As)\\
\hline
\hline
\end{tabular}
\end{center}
\end{table}

In order to check the stability of CAs$_3$ monolayer, we have calculated the binding energy (per atom) of the system in both planar and buckled configurations by using the expression  
\begin{eqnarray}
E_B =  [E_{C_2As_6}-2E_C-6E_{As}]/8
\end{eqnarray}
where $E_{C_2As_6}$, $E_C$ and $E_{As}$ represent the energies of CAs$_3$ monolayer (unit cell contains 2 carbon and 6 arsenic atoms), carbon and arsenic atoms respectively.  We find that CAs$_3$ monolayer in both planar and buckled configurations forms a bound system since the binding energies are negative in both the cases. The binding energy (per atom) of the buckled configuration is 326 meV lower than that of planar configuration. This is possibly due to the fact that the arsenic atom, being pentavalent, prefers sp$^3$ hybridization rather than sp$^2$.  

\begin{figure}[!t]
\begin{center}
\includegraphics[width=0.5\textwidth]{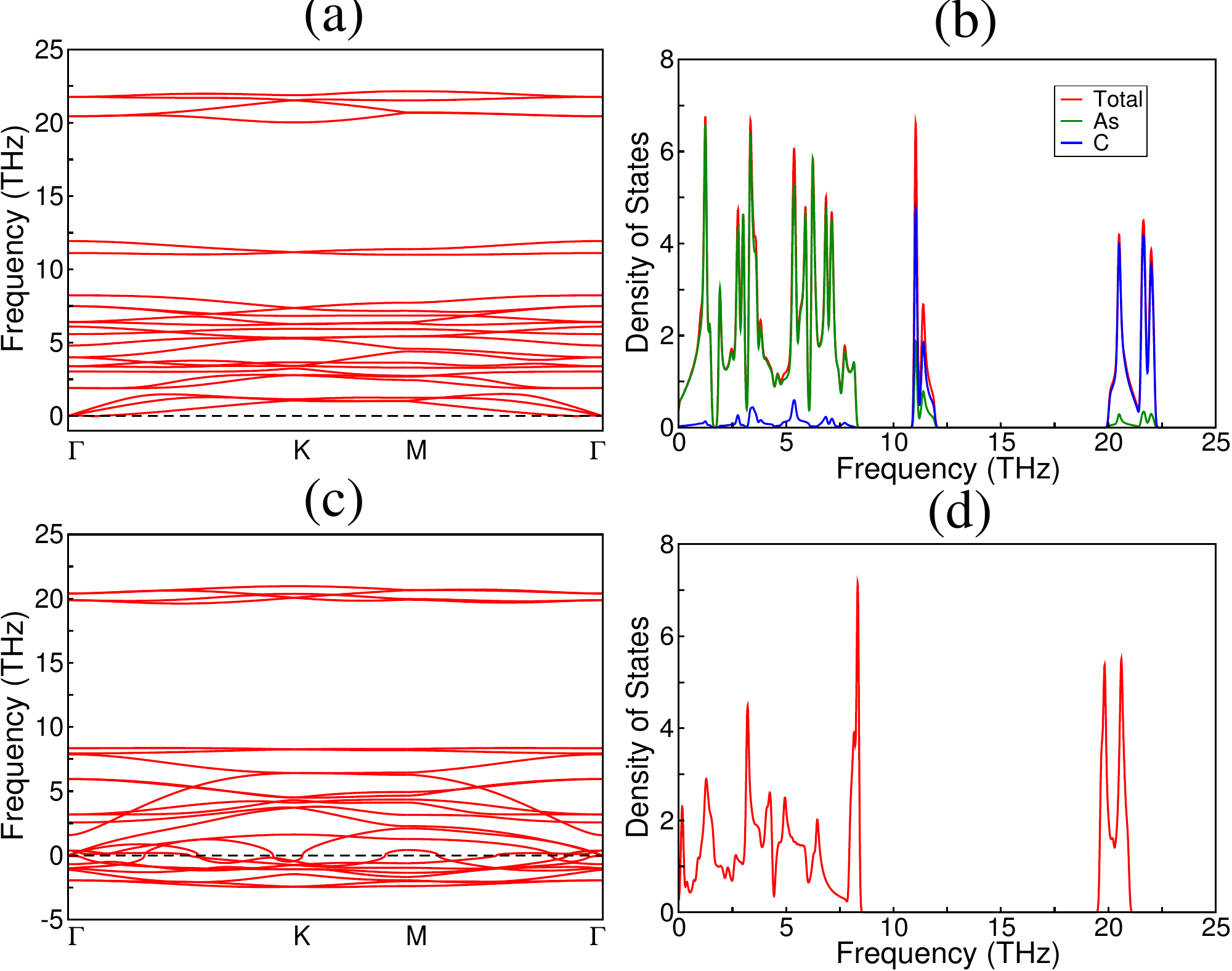}
\end{center}
\caption{ (color online) The phonon dispersion curves  along the highly symmetric k-points in Brillouin zone and  density of states for CAs$_3$ monolayer in  buckled (a and b) and planar (c and d) configurations.}
\label{FigStruct}
\end{figure}

In addition to the binding energy calculations, we have carried out phonon calculations for both  buckled and planar configurations to check the dynamical stability of CAs$_3$ monolayer. The plots for the phonon dispersions and density of states are given in Fig. 2.  The results of phonon dispersions suggest that the buckled structure is dynamically stable since the frequencies of all the 24 modes are positive. However, in case of planar configuration, we find that many of the modes possess imaginary frequencies (shown as negative frequencies)  throughout the Brillouin zone and hence the planar structure will not be dynamically stable,  though it has negative binding energy. Thus, we shall concentrate now on the properties of buckled structure and not those of dynamically unstable planar configuration. From Fig. 2(a), we observe that the phonon dispersion curves of buckled structure contain three acoustic modes in which two are longitudinal and one is transverse. Around $\Gamma$ point, the dispersion curves of the longitudinal modes vary in linear fashion whereas the transverse acoustic (ZA) mode shows a parabolic dispersion. The parabolic-like variation in transverse acoustic mode  is the characteristic of a two-dimensional system and similar variation is observed for graphene, silicene, germanene, arsenene, etc\cite{sili1,germ1,phos1,arse1}. Phonon density of states (DOS) contains four regions of vibrational energy bands (three optical and one acoustic) and three gaps exist between them (See Fig 2(b)). The smallest energy gap of 0.388 THz is observed between acoustic and first optical band. Detailed analysis of partial DOS indicates that the major contribution for the acoustic and first optical bands is due to the vibration of As atoms. The narrow band which occurs around 11 THz has contributions from both C and As atoms. On the other hand, the states with higher vibrational energy (above 20 THz) have contribution mainly due to carbon atoms. These trends are consistent with the fact that the vibrational energies of lighter C atoms is higher than those of heavier As atoms. 

\subsection{Electronic Band Structure}

\begin{figure}[!t]
\begin{center}
\includegraphics[width=0.45\textwidth]{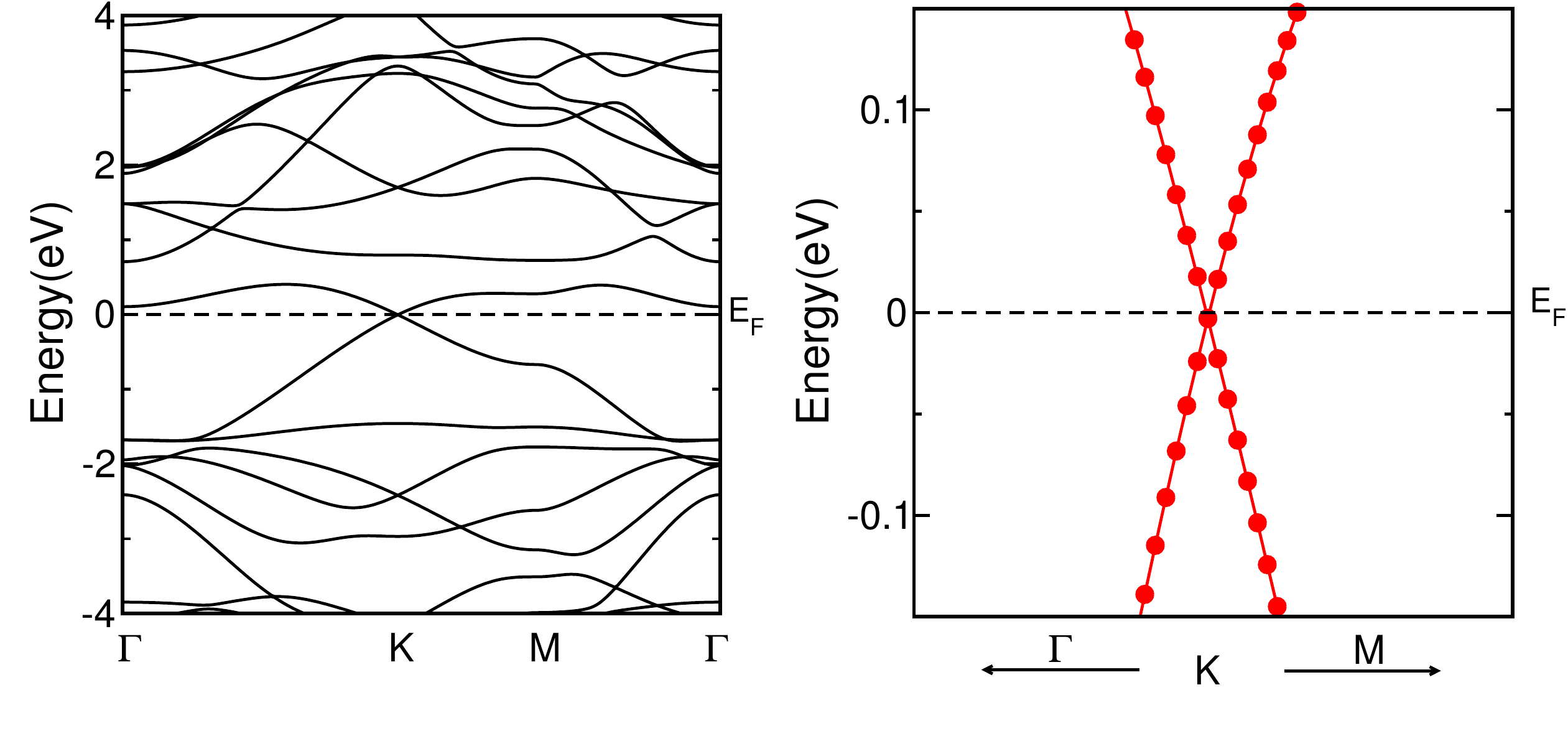}
\end{center}
\caption{ (color online) The electronic band structure of CAs$_3$ (in buckled configuration) along the highly symmetric k-points in two different energy ranges.}
\label{FigStruct}
\end{figure}

In this sub-section, we shall discuss the electronic properties of CAs$_3$ monolayer in the buckled configuration. The results for the electronic band structures (2D and 3D plots) and DOS are given in Fig. 3, 4 and 5 respectively. Results of DOS indicate that,  the monolayer is a semi-metal as the value of DOS at the E$_F$ is zero.  Most important observation from the present calculations is that the electronic band structure of CAs$_3$ monolayer contains linear dispersion around the E$_F$ at the high symmetry $K$ point. Thus, the carriers in the states close to the E$_F$ behave as massless Dirac-Fermions. These results suggest that the electronic properties of CAs$_3$ monolayer in this respect is  similar to those of graphene and other group IV monolayers. The range of linear dispersion in the valence and conduction bands is from -600 to 250 meV. However, in the conduction region, there is a parabolic-like dispersion curve  corresponding to Schrodinger-like particles at $\Gamma$ point around 105 meV.  In Fig. 4, we have also drawn the 3D plots for (a) highest occupied conduction and (b) lowest unoccupied valence bands in their full energy range.  The sub-figures (c) and (d) show the combined bands in two different energy ranges. The top and bottom of xy plane also contain contour plots of the bands. We can clearly see the presence of Dirac cones in the 3D plot of energy versus (k$_x$, k$_y$). There exist six Dirac cones at the edges of the hexagons (high symmetry $K$ point) around the E$_F$. The presence of paraboloid-like band with minima at $\Gamma$ point  is also visible in sub-figures (a) and (d). The bottom of the paraboloid is at 105 meV above the E$_F$. Due to this, the range of observing purely Dirac-Fermion like behaviour of particle (electron-like) will be restricted from 0 to 105 meV.  For any excitations with energy above 105 meV, we shall observe a mixed behaviour due to  the presence of both  Dirac and Schrodinger like Fermions. We would like to mention here that the energy of 105 meV is much higher than the thermal energy k$_B$T at the room temperature. Thus, most of the charge carriers in this material are expected to behave as Dirac-Fermions at the room temperature. Similar to graphene, it may be possible to tune the position of the E$_F$ from 105 meV to -600 meV by appropriate gate voltage and thus, increase or decrease the value of conductivity of Dirac-Fermion. Furthermore, the polarity of the carriers (electron or hole like) will be decided by the position of the E$_F$.  It may also be  possible to tune the E$_F$ by external influences like,  electron or hole doping, strain, etc. We shall discuss the effect of mechanical strain on the properties of CAs$_3$ monolayer in the next sub-section.
\begin{figure}[!t]
\begin{center}
\includegraphics[width=0.5\textwidth]{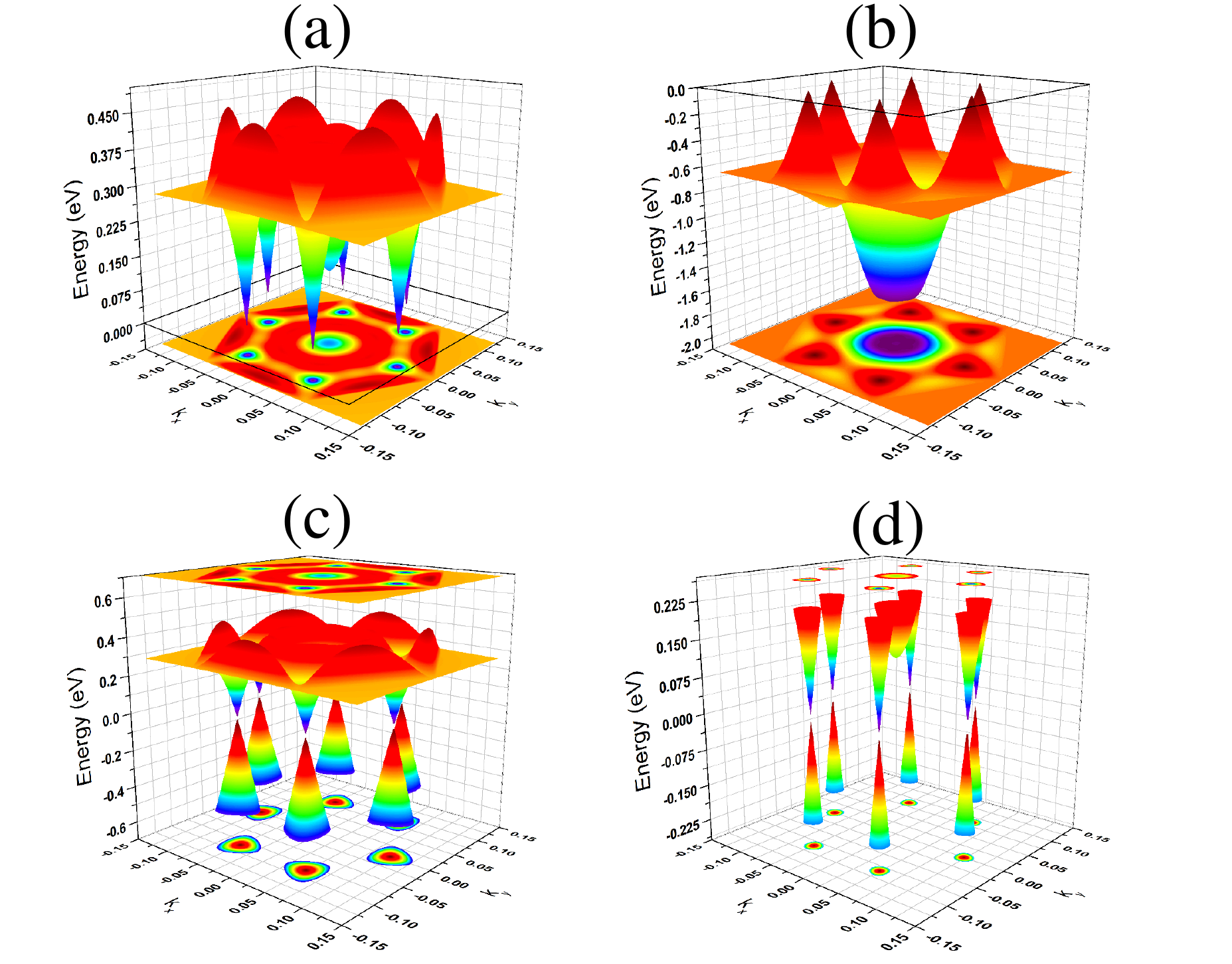}
\end{center}
\caption{ (color online) Three-dimensional (3D) plots of electronic band structure of CAs$_3$ monolayer in buckled configuration: (a) lowest unoccupied conduction and (b) highest occupied valence  bands which are close to the Fermi level. Combined plots of these two  bands are given (c) and (d) in two different energy ranges. Presence of Dirac cones at the Fermi level is clearly visible in (d).}
\label{FigStruct}
\end{figure}

In order to understand the nature of states around the E$_F$, we have also carried out detailed analysis of the partial electronic DOS (See Fig. 5). 
Our analysis suggests that both the valence and conduction bands (Fig. 5) have contributions from  s and p orbitals of both the constituent atoms. It clearly indicates the presence of strong hybridization between the orbitals of As and C atoms. By analyzing further, we find that the contributions from 3p orbitals of As atom is predominant in the range of energies presented in Fig. 5, except for the states close to the E$_F$. This is due to fact that the unit cell contains more number of As atoms which have more number of electrons in 3p orbitals as  compared to C atoms (2p orbitals). In case of states around the E$_F$ (from -1.5 to 0.5 eV), it is clearly visible from partial DOS that the amount of contribution of 2p orbitals of C atom is at par with that of 3p orbitals of As atoms. As mentioned above the number of C atoms in the unit cell is two as compared to six As atom. Thus, if we scale the partial DOS of C and As atom by their numbers in the unit cell (2 and 6 respectively), the states close to the E$_F$ have more 2p orbital (of C atom) character as compared to 3p orbital of As atom. This observation is quite important because it gives satisfactory explanation for the presence of linear dispersion around the E$_F$. 

\begin{figure}[!t]
\begin{center}
\includegraphics[width=0.45\textwidth]{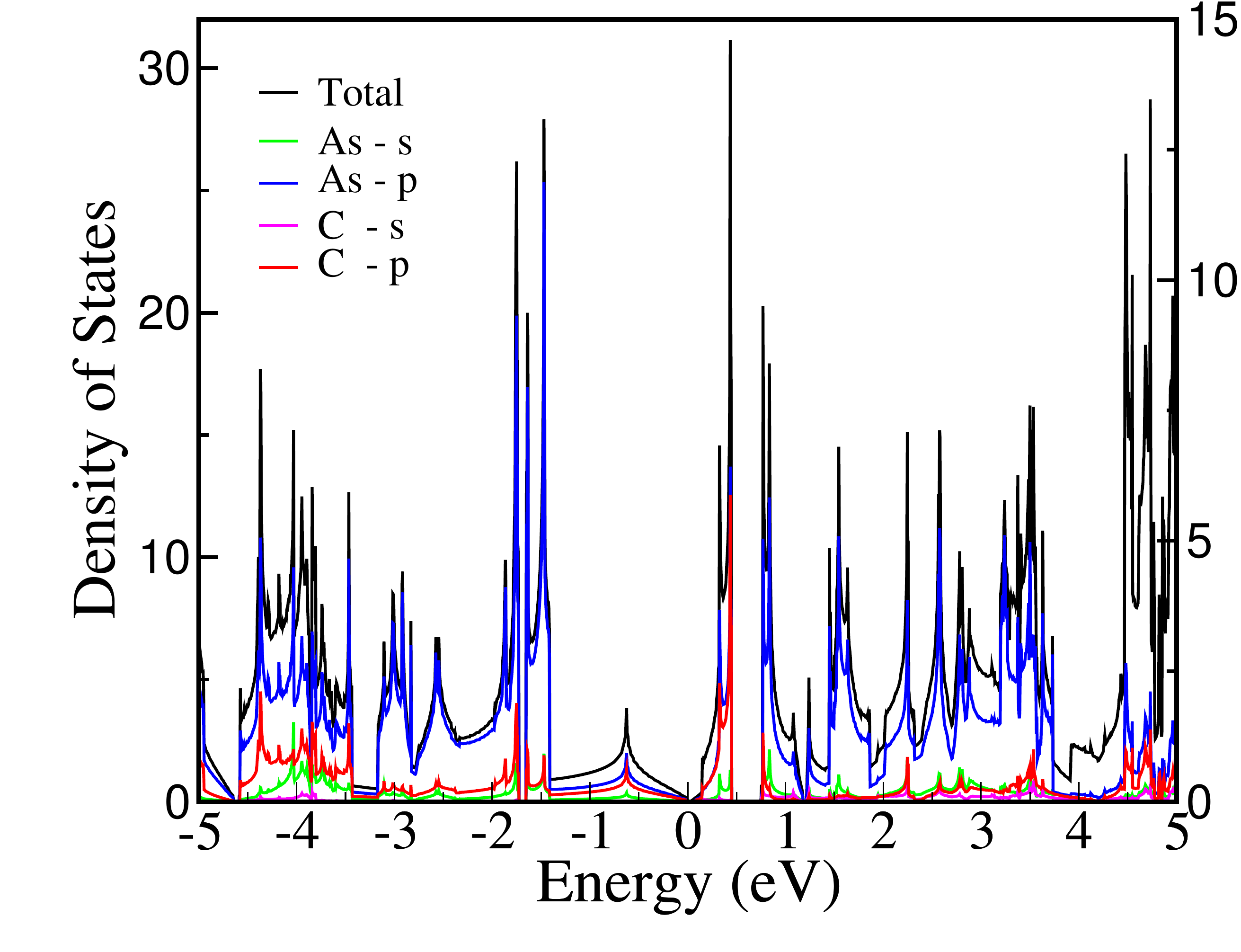}
\end{center}
\caption{ (color online) Total (DOS) and partial (PDOS) electronic density of states  for CAs$_3$ monolayer in buckled configuration.}
\label{FigStruct}
\end{figure}

It will be interesting to compare the properties of CAs$_3$ monolayer with those of buckled arsenene, because the former can be thought as carbon substituted arsenene. In buckled arsenene monolayer,  As atom,  with five valence electrons, favors sp$^3$ hybridization (having a molecular analog of NH$_3$), where four orbitals (3s, 3p$_x$, 3p$_y$ and 3p$_z$) of As atom are combined to produce four hybridized orbitals. Among these four orbitals, three orbitals form three covalent bonding (3 electrons from each As atom) with its three nearest neighbors and the remaining two electrons act as a lone pair which does not contribute to the bonding. This makes arsenene monolayer a semi-conductor.  On the other hand, the situation in CAs$_3$ monolayer is slightly different because of the two C atoms present in the unit cell. The results of geometric structure for the buckled configuration show that the bond angles around C atom is 116.09$^\circ$ which deviates slightly from 120$^\circ$ (characteristic of sp$^2$ hybridization), but it is far away from 109.47$^\circ$ (characteristic of sp$^3$ hybridization). Thus, C atoms in CAs$_3$ monolayer prefers sp$^2$-like hybridization in which one s and two p orbitals (2p$_x$ and 2p$_y$) combine to produce three hybridized orbitals which in turn make three covalent-like bonds (three valence electrons of C) with its three nearest neighbor As atoms (one electron each). The fourth electron will be in the unhybridized p$_z$ orbital. However, it is important to note that since the bond angles centered around C atoms is not perfectly 120$^\circ$, the hybridization cannot purely sp$^2$. Thus, it is expected that there will be a small mixing between the 2p$_z$ and sp$^2$ hybridized orbitals. Our analysis of partial DOS indicates that the valence and conduction bands around the E$_F$ are due to these mixed orbitals.
\begin{figure}[!t]
\begin{center}
\includegraphics[width=0.45\textwidth]{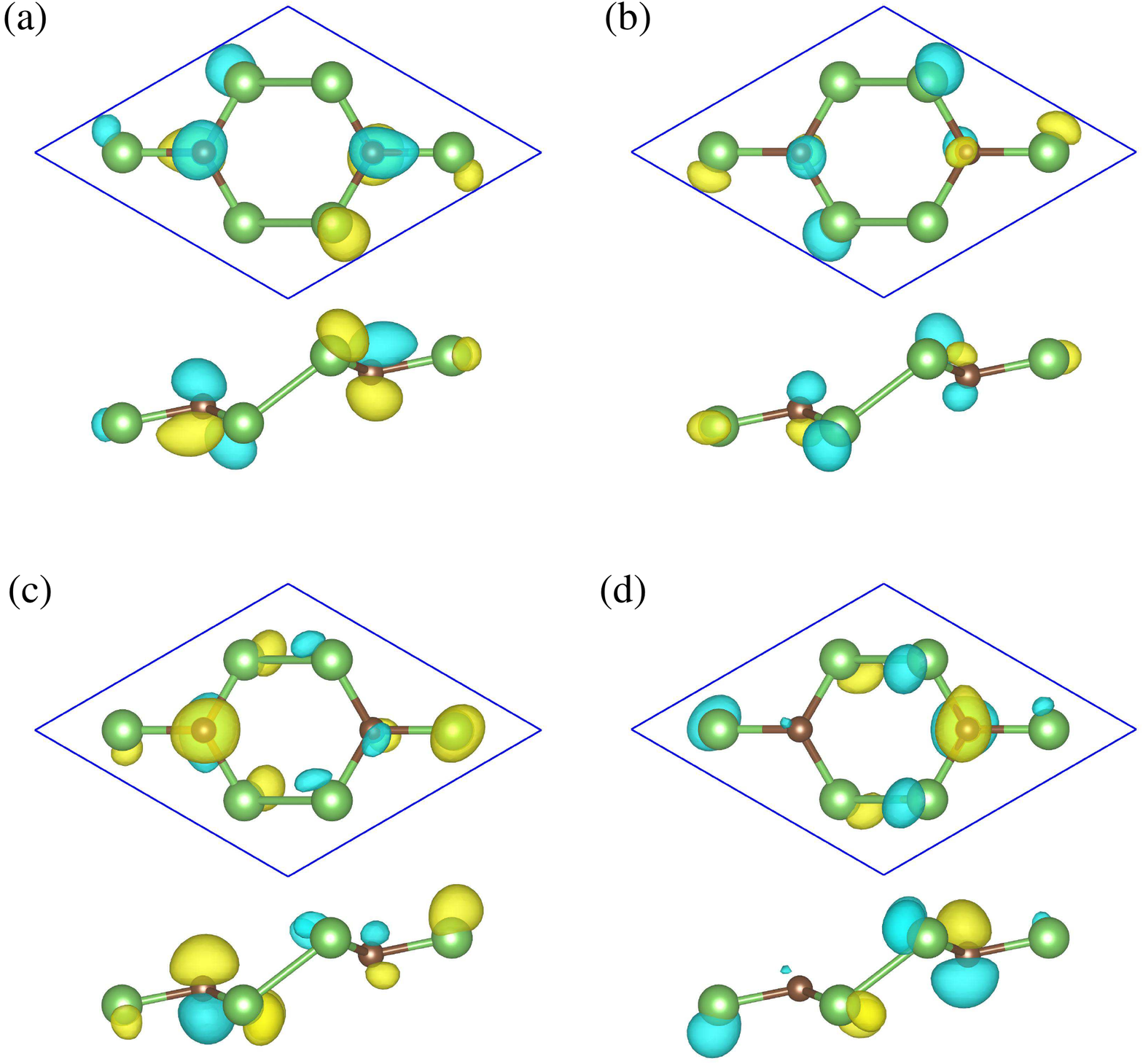}
\end{center}
\caption{ (color online) The spatial distributions of Kohn-Sham states at Dirac point which lies at the high symmetric $K$ and the E$_F$. Real and imaginary parts of occupied (a and b) and unoccupied (c and d) orbitals. }
\label{FigStruct}
\end{figure}
In order to verify this fact, we have plotted the Kohn-Sham (KS) states corresponding to highest occupied valence and lowest unoccupied conduction states in Fig. 6. It is clear from the figure that these two states are degenerate and occur at the high symmetric $K$ point in the Brillouin zone. The plots of KS states clearly show the presence of  two-lobed (dumbbell-like shape) of 2p$_z$ around C atoms. We can also see a contribution from orbitals of As atoms due to the above mentioned mixing between p$_z$ and sp$^2$ hybridized orbitals. This establishes the fact that 2p$_z$ orbital of C atom plays an important role in determining the nature of dispersions, in the present case,  linear dispersion (Dirac cone) around the E$_F$.  It is important to note here that the bond angle between the Si atoms in silicene\cite{sili-ck1} is 116.08$^\circ$ which is quite close to the angle As-C-As observed in the present case.  This indicates the similarity between the CAs$_3$ monolayer and silicene where the hybridization is a mixture of sp$^2$ and 3p$_z$ orbitals of Si atom. Further, in case of silicene, the overlap between the 3p$_z$ orbitals is weak due to the large internuclear distance Si-Si (as compared to C-C in graphene).  We have also plotted the valence electron charge density distribution for CAs$_3$ monolayer in three different isovalues (0.03, 0.07 and 0.11 e/Bohr$^3$) in Fig. 7. For lower isovalue of density,  the distribution of charges around all the constituent atoms is almost uniform. However, the distribution of charge density in Fig. 7 (b) shows the presence of more overlapping region between As and C atoms than those between As atoms themselves.  On the other hand, more accumulation of charges around C atoms is observed for higher isovalue of density.

\begin{figure}[!t]
\begin{center}
\includegraphics[width=0.5\textwidth]{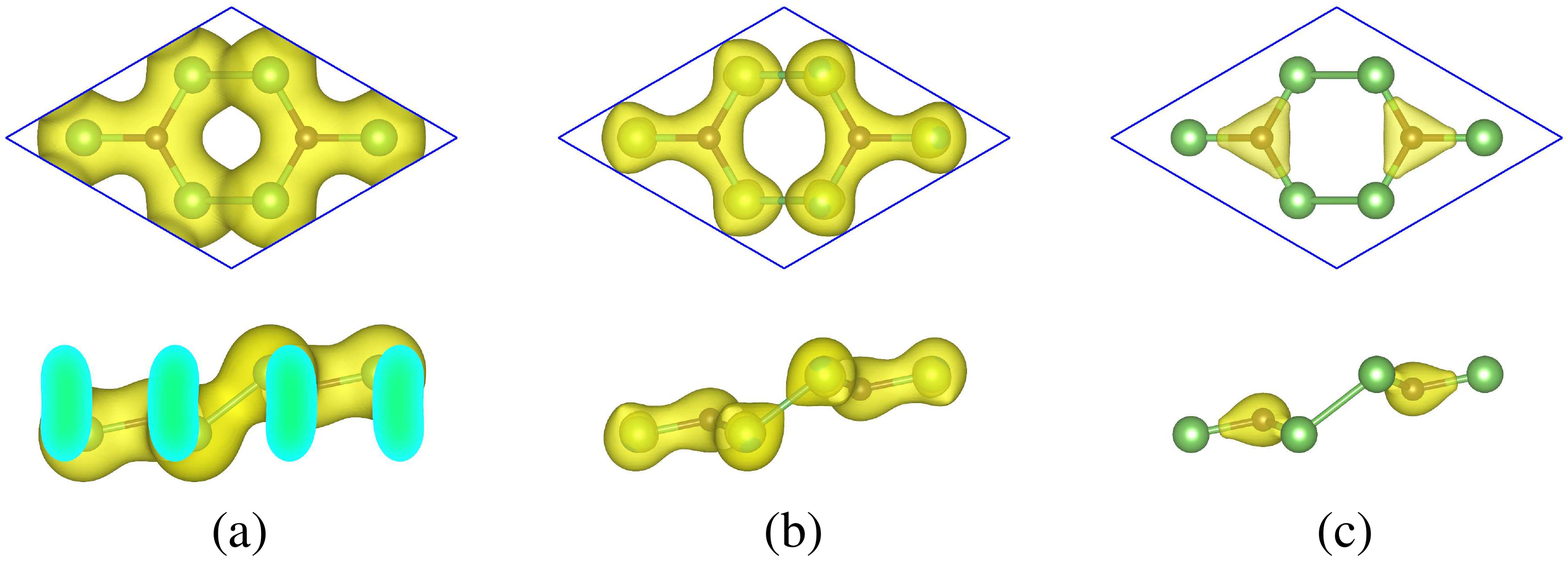}
\end{center}
\caption{ (color online) The spatial distribution of  valence charge density for CAs$_3$ monolayer in different isovalues (0.03, 0.07 and 0.11 e/Bohr$^3$).}
\label{FigStruct}
\end{figure}

\subsection{Effect of Mechanical Strain}

\begin{figure}[!t]
\begin{center}
\includegraphics[width=0.45\textwidth]{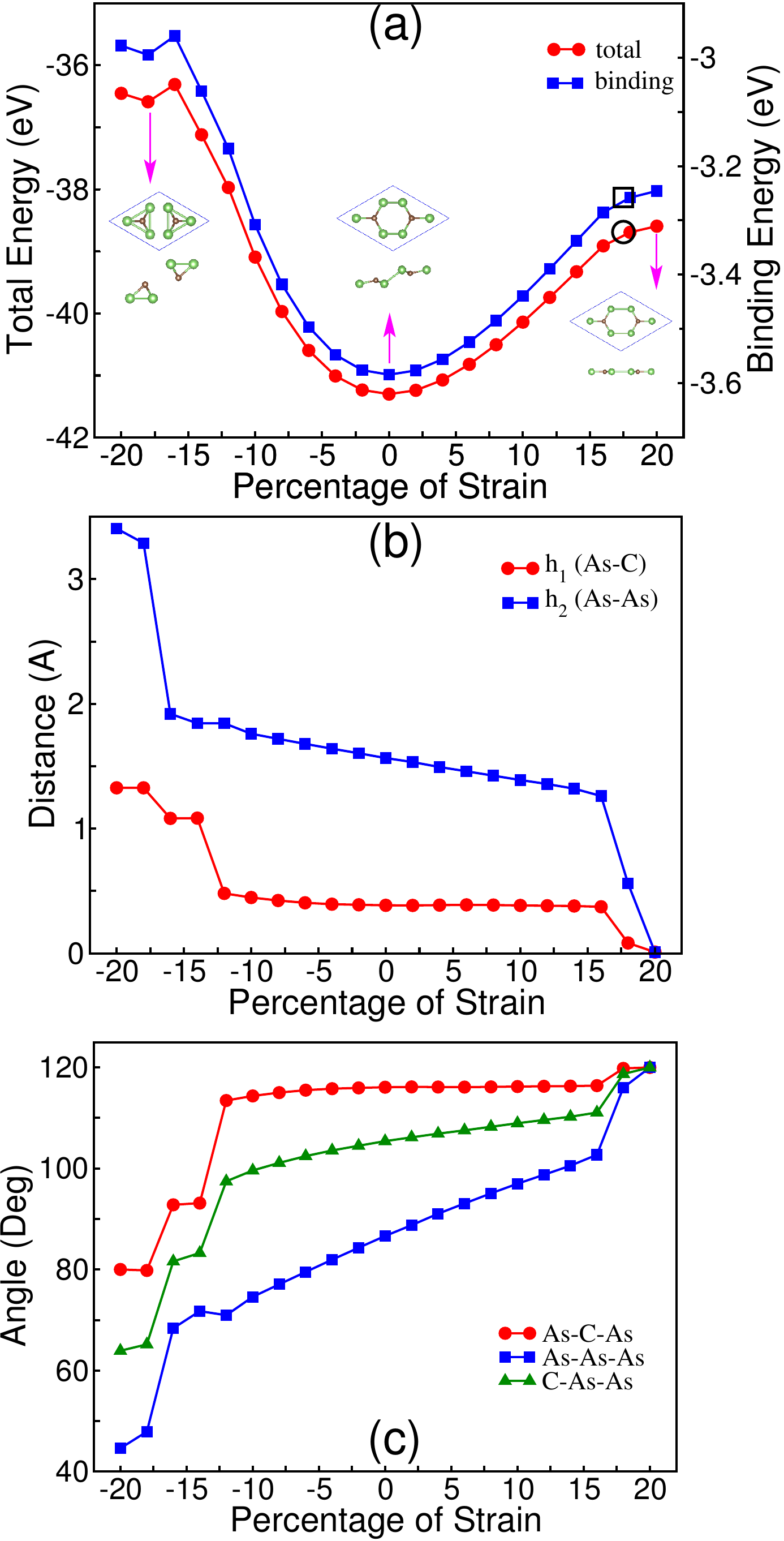}
\end{center}
\caption{ (color online) Variation of (a) total and binding energies, (b) distances (h$_1$, h$_2$) and (c) angles  for CAs$_3$ monolayer  with the bi-axial strain. The open square and circle in (a) represent the total and binding energies of the planar configuration.}
\label{FigStruct}
\end{figure}

Having discussed the electronic properties of CAs$_3$ monolayer in previous sub-section, now we shall focus our attention on the results of study of  effect of mechanical strain on the geometrical and electronic properties of CAs$_3$ monolayer. 

To study the influence of mechanical strain,  we have performed the electronic structure calculations of CAs$_3$ monolayer under bi-axial, both compressive and tensile, strains.  The effect of strain has been simulated by freezing the lattice constant $a$ to a particular value other than the equilibrium lattice constant $a_0$ (6.778 $\AA$). With this fixed value of lattice constant, the fractional coordinates of all the atoms in the unit cell have been optimized with the convergence criteria mentioned in Section II. The percentage of strain is calculated by using the expression $(a-a_0)/a_0$. The positive and negative values of strain represent the tensile and compressive strains, respectively. In  the present study, we have varied the values of strains from -20\% to 20\% in steps of 2\%.

\begin{figure*}[!t]
\begin{center}
\includegraphics[width=1.0\textwidth]{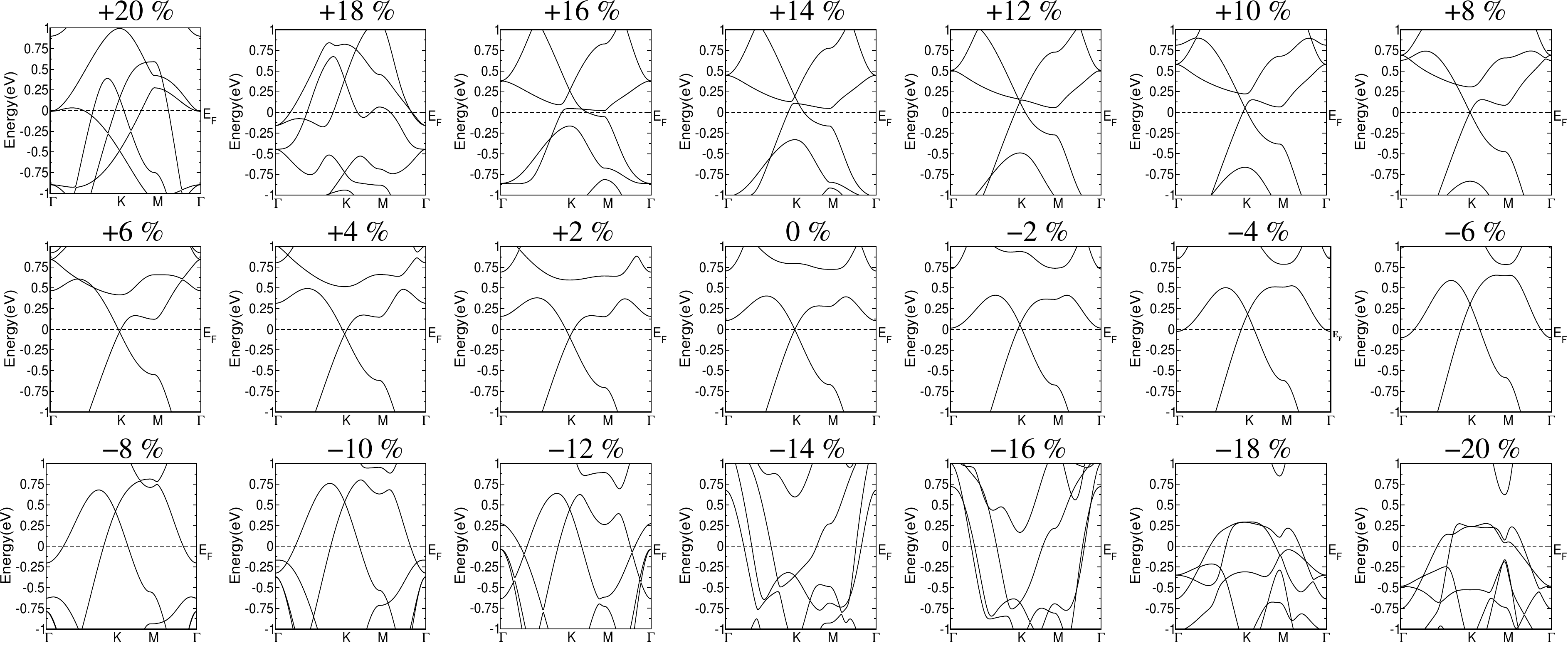}
\end{center}
\caption{ (color online) Variation of electronic band structure of CAs$_3$ monolayer with bi-axial strains from -20\% to +20\%.}
\label{FigStruct}
\end{figure*}

\subsubsection{Geometry and Energetics}
Our results for the energetics (total and binding energies), geometrical parameters (buckling lengths, bond angles) and electronic band structures of CAs$_3$ monolayer with mechanical strains are summarized in Fig. 8 and 9. Variation of total and binding energies with strain shows a  parabolic curve with a stable minima at the equilibrium lattice constant (See Fig. 8 (a)). However, when the magnitude of the strain is large, there is a clear deviation from the parabolic trend.  Further, it is observed that both the buckling lengths h$_1$ and h$_2$ change monotonically with the strain, though the rate of variation in h$_2$ is larger as compared to that in h$_1$.  When the tensile strain reaches 18\%, the values of these two buckling lengths decrease sharply and become nearly zero ($\sim$ 0.01 $\AA$) at 20\% of strain. Furthermore,  the two bond angles namely, As-As-As and C-As-As,  also show a clear monotonic rise with increasing the tensile strain. In case of the angle As-C-As, we observe that the slope of the variation is very small.  In this case, angle is increased from 116.09$^\circ$ to 116.40$^\circ$ when we go from equilibrium structure to 16\% of tensile strain. All the three angles suddenly decrease around 18\% of strain and reach almost 120$^\circ$ at 20\% of strain. This indicates that the buckled configuration is reduced to the planar configuration. For the sake of comparison, the data points corresponding to the total and binding energies of the planar configuration are included (black colored open symbols) in Fig. 8 (a). These points fall closely to those of the buckled configuration with 18\% of strain. It is to be noted here that the value of lattice constant for the planar configuration (7.964 $\AA$) is equal to that of buckled configuration with 17.5\% of tensile strain. On the other hand, for compressive strains, the value of h$_1$ increases slowly until there is a sharp increase at -14\% of strain. Similarly, the value of h$_2$ also undergoes a drastic rise at -18\% of strain. Further, we also observe abrupt changes in the values of three angles beyond -14\% of compressive stains. These results clearly indicate that the bonds between As atoms are broken and the geometry of atoms in the unit cell turns as two molecular units of CAs$_3$ (See insert in Fig. 8(a) for geometry with strain of -18\%).  

\subsubsection{Electronic Structure}
In Fig. 9, we plot the electronic band structure of buckled CAs$_3$ monolayer with different values of mechanical strain, ranging from -20\% to +20\%. Many important modifications in the electronic band structures due to the application of mechanical strain on the system have been observed. Firstly, it is clearly seen that the linear dispersion disappears for compressive strain beyond -12\%. The presence of  the linear dispersion (and hence Dirac cone) around the high symmetric $K$ point is clearly observed for an applied strain ranging from -12\% to +20\%. However, the Dirac point may not lie exactly at the E$_F$. Reason for the disappearance of linear dispersion beyond -12\% can be traced back to the modification in the geometry and hence the hybridization. When the compressive strain change from  -12\% to -14\%, there are drastic changes in the values of bond angles (As-C-As, C-As-As) and buckling length (h$_1$, which is a vertical length between  C and As atoms). This clearly indicates that major changes occur in the geometrical arrangements around C atoms. These changes will strongly influence the hybridization and hence the electronic properties of the system. For instance, the bond angle As-C-As is changed from 113.45$^\circ$ to 93.15$^\circ$ when we go from -12\% to -14\% of strain. Due to this, the overlap of 2p$_z$ orbital of C with those of sp$^2$ orbitals becomes so large that the 2p$_z$ orbital almost looses its individuality as an unhybridized orbital. In this case, the mixing leads to a situation similar to sp$^3$ hybridization. Subsequently, the feature of linear dispersion disappears in the electronic band structure. 

As we have discussed in the previous sub-section, the Dirac point lies exactly at the E$_F$ for CAs$_3$ monolayer without any strain. When CAs$_3$ monolayer is subjected to compressive strain ranging from 0 to -12\%,  it is observed that the position of the Dirac point is systematically pushed upward and it reaches maximum energy of 0.47 eV above the E$_F$. Thus, the mechanical strain can be useful tool to tune the position of the Dirac point. It is also important to note that the parabolic-like dispersion, which exists around the $\Gamma$ at 105 meV above the E$_F$ in the unstrained system, has also been pushed below the E$_F$ due to the compressive strain. It is getting occupied by the electron at -4\% of strain. Similarly, we also note that few occupied parabolic-like bands (which were much below -1.0 eV) around the $\Gamma$ points have also been pushed upward towards the E$_F$ due to the compressive strain. In case of tensile strain, we observe that the position of the Dirac point fluctuates around the E$_F$ for small amount of strain of about less than 10\% and it moves upward for a strain larger than this amount. It is also observed that the conduction bands which lie above 0.7 eV in the unstrained monolayer start moving towards the E$_F$ due to the  application of tensile strain. Importantly, there are no other dispersion curves which cross the E$_F$ or come closer to the Dirac point for the tensile strain range from 0 to 10\%. Thus, for this range of tensile strain, the transport properties of the system is mainly characterized by the massless Dirac Fermions. For 12\% of strain, we observe that one of the high lying conduction bands comes close to the Dirac point and it crosses the Dirac point for higher strain. Hence, this band may lie exactly at the Dirac point for a value of strain little above 12\%. In this situation, these three bands touching at a point may lead to the appearance of triplet Fermions in the system which is similar to one observed in $\beta_{12}$ borophene\cite{moto-prb2017}.  For strain beyond 16\%, few conduction bands cross the E$_F$ which make both Dirac- and Schrodinger-like Fermions to contribute to the electronic properties of CAs$_3$ monolayer. Overall, we find that the mechanical strain can be an useful tool to modify the properties of CAs$_3$ monolayer. 

\section{Conclusion}
Using density functional theory based electronic structure calculations, we have predicted the energy and dynamical stabilities of  CAs$_3$ monolayer in buckled configuration. The geometrical arrangement in the buckled configuration is slightly different from those observed for silicene and germanene. We have observed from the results of electronic structure calculations that CAs$_3$ monolayer is a semi-metal. In addition, our results indicate that the electronic band structure of this monolayer possesses a linear dispersion (2D band structure, E vs k) at the Fermi level around the high symmetric $K$ point of the reciprocal lattice. The presence of a Dirac cone around the $K$ point at the E$_F$ is also confirmed in the three-dimensional electronic band structure ( E vs (k$_x$, k$_y$)).  Thus, the charge carriers in this system behave as massless Dirac-Fermion for low energy excitation ( $<$  105 meV). This suggests that the properties of CAs$_3$ monolayer are similar to those of graphene, silicene and germanene. Our detailed analysis of partial density of state indicates that the 2p$_z$ orbital of C atoms plays important role in determining the presence of linear dispersion (and hence the Dirac cone) around the E$_F$.  We have also studied  the effect of mechanical strain on the properties of CAs$_3$ monolayer.  For a tensile strain beyond 18\% the buckled configuration is reduced to the planar configuration. There are drastic changes in the geometrical environment around C atom in the unit cell when the compressive strain increased beyond -12\% and hence the hybridization between the orbitals of C atoms changes. This leads to the disappearance of linear dispersion in the electronic band structure. We also have observed a systematic upward shift in the position of the Dirac point due to the compressive strain which can be an useful tool to the modify the properties of CAs$_3$ monolayer. Finally, it is important to note that this monolayer system belongs to the class of \textit{Dirac materials} and hence, like graphene,  several properties of this system at low energy excitation are governed by the Dirac-like Hamiltonian. 

\section{Acknowledgment}
The author thanks Dr. Aparna Chakrabarti for critical reading of the manuscript and fruitful discussions.  The author also thanks Dr. Arup Banerjee for the discussions.  Dr. P. A. Naik is thanked for constant support and encouragement.

% \newpage


\begin{thebibliography}{99}
\bibitem{graphene} A. H. Castro Neto, F. Guinea, N. M. R. Peres, K. S. Novoselov, and A. K. Geim, Rev. Mod. Phys., {\bf 81}, 109 (2009) and references therein.
 \bibitem{graphene1} M. I. Katsnelson, Graphene: Carbon in Two Dimensions, (Cambridge Univ. Press, Cambridge, 2012). 
\bibitem{boro1} J. Kunstmann, A. Quandt, Phys. Rev. B {\bf 74}, 035413 (2006)
\bibitem{boro2} H.  Tang,  S.  I. Beigi, Phys. Rev. Lett. {\bf 99}, 115501 (2007)
\bibitem{boro3}  X.  Wu, J.  Dai, Y. Zhao, Z. Zhuo, J. Yang, X. C. Zeng, ACS Nano, {\bf6}, 7443 (2012)
\bibitem{alum1} C. Kamal, A. Chakrabarti and M. Ezawa, New J. Phys., {\bf 17},  083014 (2015).
\bibitem{sili1} K. Takeda, K. ShiraishiPhys. Rev. B {\bf 50}, 14916 (1994)
\bibitem{sili2} G. G. Guzmán-Verri and L. C. Lew Yan Voon, Phys. Rev. B {\bf 76}, 075131 (2007)
\bibitem{sili3} S. Cahangirov, M. Topsakal, E. Aktürk, H. Şahin, and S. Ciraci, Phys. Rev. Lett. {\bf 102}, 236804 (2009)
\bibitem{sili-ck1} C. Kamal, A. Chakrabarti, A. Banerjee and S. K. Deb, J. Phys. Cond. Mat., {\bf 25}, 085508 (2013).
\bibitem{sili-ck2} C. Kamal, A. Banerjee, A. Chakrabarti, Chapter 15, Pages 221-234, {\it Graphene Science Handbook: Size-Dependent 
Properties}, Edited by M Aliofkhazraei,{\it et al.}, CRC Press (2016) and references therein. 
\bibitem{sili-ck3}  C. Kamal, A. Chakrabarti, A. Banerjee, Phys. Lett. A, \textbf{378}, 1162 (2014)
 \bibitem{group4_1} C. C. Liu, H. Jiang, and Y. Yao, Phys. Rev. B, {\bf 84}, 195430 (2011).
\bibitem{group4_2} J. C. Garcia, D. B. de Lima, Lucy V. C. Assali, J. F. Justo, J. Phys. Chem. C, {\bf 115}, 13242 (2011)
\bibitem{germ1}  A. Acun, L. Zhang, P. Bampoulis, M. Farmanbar, A. van Houselt, A. N. Rudenko, M. Lingenfelder, G. Brocks, B. Poelsema, M. I. Katsnelson and H. J. W. Zandvliet, J. Phys. Cond. Mat., {\bf 27}, 443002 (2015).
\bibitem{sn1}  B. van den Broek, M. Houssa, E. Scalise, G. Pourtois, V. V. Afanasev and A. Stesmans, 2D Materials, {\bf 1}, 021004 (2014).
\bibitem{plum1} P. Rivero, J. A. Yan, V. M. García-Suárez, J. Ferrer, S. Barraza-Lopez, Phys. Rev. B {\bf 90}, 241408(R) (2014).
\bibitem{nitr1} V. O. Ozcelik, O. U. Akturk, E. Durgun, S. Ciraci, Phys. Rev. B \textbf{92} 125420 (2015)
\bibitem {phos1} J. Qiao, X. Kong, Z. X. Hu, F. Yang, W. Ji, Nat. Comm., {\bf 5}, 4475 (2014). 
\bibitem {phos2} V. Tran, R. Soklaski, Y. Liang and L. Yang, Phys. Rev. B, {\bf 89}, 235319 (2014).
\bibitem {phos3} A. N. Rudenko and M. I. Katsnelson, Phys. Rev. B {\bf 89}, 201408(2014).
\bibitem{arse1} C. Kamal, M. Ezawa, Phys. Rev. B, {\bf 91}, 085423 (2015).
\bibitem{anti1} S. Zhang, Z. Yan, Y. Li, Z. Chen, H. Zeng, Angew. Chem. Int. Ed. {\bf 54}, 3112 (2015).
\bibitem{bism1} E. Akt\"urk, O. Uzengi Akt\"urk, and S. Ciraci, Phys. Rev. B, {\bf 94}, 014115 (2016)
\bibitem{Eboro1}  A. J. Mannix, X. F. Zhou, B. Kiraly, J. D. Wood, D. Alducin, B. D. Myers, X. Liu, B. L. Fisher, U. Santiago, J. R. Guest,  M. J. Yacaman, A. Ponce, A. R. Oganov, M. C. Hersam, N. P. Guisinger, Science, {\bf 350}, 1513 (2015).
\bibitem{Eboro2} B. Feng, J. Zhang, Q. Zhong, W. Li, S. Li, H. Li, P. Cheng, S. Meng, L. Chen, K. Wu, Nature Chemistry {\bf 8}, 563 (2016)
\bibitem{Esi1} B. Lalmi, H. Oughaddou, H. Enriquez,  A. Kara, S. Vizzini, B. Ealet, B. Aufray, Appl. Phys. Lett. {\bf 97}, 223109 (2010)
 \bibitem{Esi2}P. Vogt, P. D. Padova, C. Quaresima, J. Avila,  E. Frantzeskakis, M. C. Asensio, A. Resta, B. Ealet, L. G. Lay, Phys. Rev. Lett., {\bf 108},155501 (2012).
\bibitem{Esi3} A. Fleurence, R. Friedlein, T. Ozaki, H. Kawai, Y. Wang, and Y.  Y. Takamura, Phys. Rev. Lett., {\bf 108},245501 (2012).
\bibitem{Esi4} L. Tao, E. Cinquanta, D. Chiappe, C. Grazianetti, M. Fanciulli, M. Dubey, A. Molle, D. Akinwande, Nature Nanotechnology {\bf 10} 227 (2015).
\bibitem{Ege1} M. E. Davila, L. Xian, S. Cahangirov, A. Rubio, G. L. Lay, New J. Phys. {\bf 16} 095002 (2014)
\bibitem{Ege2} M. Derivaz,  D. Dentel, R. Stephan, M-C. Hanf, A. Mehdaoui, P. Sonnet, C. Pirri, Nano Lett.{\bf 15}, 2510 (2015).
\bibitem{Ege3} M. E. Davila, G. L. Lay,  Scientific Reports {\bf 6}, 20714 (2016)
\bibitem{Esn1} F. F. Zhu, W. J. Chen, Y. Xu, Chun-lei Gao, Dan-dan Guan, Can-hua Liu, Dong Qian, Shou-cheng Zhang, Jin-feng Jia, Nat. Mater., {\bf 14}, 1020 (2015).
\bibitem{Esn2} S. Saxena, R. P. Chaudhary, S. Shukla, Sci. Rep. {\bf 6}, 31073 (2016) 
\bibitem{Esn3} J. Gao, G. Zhang, Y. W. Zhang, Sci. Rep. {\bf 6},  29107 (2016)
\bibitem{Eph1} L. Li, Y. Yu, G. J. Ye, Q. Ge, X. Ou, H. Wu, D. Feng, X. H. Chen, and Y. Zhang, Nat. Nanotech., {\bf 9}, 372 (2014).
\bibitem{Eph2}  A. Castellanos-Gomez, L. Vicarelli, E. Prada, J. O. Island, K. L. Narasimha-Acharya, S. I. Blanter, D. J. Groenendijk, M. Buscema, G. A. Steele, J. V. Alvarez, H. W. Zandbergen, J. J. Palacios, H. S. J. van der Zant, 2D Materials {\bf 1}, 025001 (2014).
\bibitem{Eas1}H. S. Tsai, S. W. Wang, C. H. Hsiao, C. W. Chen, H. Ouyang, Y. L. Chueh, H. C. Kuo, J. H. Liang, Chem. Mater., {\bf 28} 425 ( 2016)
\bibitem {Egroup5_1} M. Pumera and Z. Sofer, Adv. Mater., {\bf 29}, 1605299 (2017).
\bibitem {Egroup5_2} F. Reis, G. Li, L. Dudy, M. Bauernfeind, S. Glass, W. Hanke, R. Thomale, J. Schäfer, R. Claessen, Science, {\bf357} 287 (2017)
\bibitem{sili-gap1} Z. Ni, Q. Liu, K. Tang, J. Zheng, J. Zhou, R. Qin, Z. Gao, D. Yu, J. Lu,  Nano Lett., {\bf 12}, 113(2012).
\bibitem{sili-gap2} N. D. Drummond, V. Z\'olyomi, V. I. Falko, Phys. Rev.B, {\b 85}, 075423(2012).
\bibitem{sili-gap3} M. Ezawa,  New J. Phys., {\bf 14},  033003 (2012). 
\bibitem {sili-gap4} C. Kamal, arXiv:1202.2636v1[cond-mat.mes-hall],(2012).
\bibitem{dirac} T.O. Wehling, A.M. Black-Schaffer, A.V. Balatsky, Adv. in Physics, {\bf 63}, 1-76 (2014).
\bibitem {group3-5} H. Sahin, S. Cahangirov, M. Topsakal, E. Bekaroglu, E. Akturk, R. T. Senger, S. Ciraci, Phys. Rev. B {\bf80}, 155453 (2009)
\bibitem{group-4-6} C. Kamal, A. Chakrabarti, M. Ezawa, Phys. Rev. B, {\bf 93}, 125428 (2016).
\bibitem {gep3} Y. Jing,  Y. Ma, Y. Li, T. Heine, Nano Lett., {\bf17} 1833 (2017). 
\bibitem {snp3-1} B. Ghosh, S. Puri, A. Agarwal, S. Bhowmick, J. Phys. Chem. C, {\bf 122} 18185 (2018)
\bibitem {snp3-2} L. P. Feng, A. Li, P. C. Wang, Z. T. Liu, J. Phys. Chem. C, {\bf 122} 248539 (2018)
\bibitem {flex-2d} L. T. Hoa, H. N. Tien, V. H. Luan, J. S. Chung, S. H. Hur, Sensors and Actuators B: Chemical {\bf 185}, 701 (2013); Z. L. Wang, Appl. Phys. A, {\bf 88}, 7 (2017); P. K. Kanna, D. J. Late, H. Morgan, C. S. Rout, Nanoscale, {\bf 7}, 13293 (2015).
\bibitem{vasp} G. Kresse, J. Furthmuller, Phys. Rev. B {\bf 54}, 11169 (1996); 
G. Kresse, D. Joubert, Phys. Rev. B {\bf 59}, 1758 (1999); VASP 5.2 programme package used here is fully integrated in the MedeA platform (Materials Design, Inc.) with a  graphical user interface for the computation of the properties.
\bibitem{paw} P. E. Bl\"ochl, Phys. Rev. B, {\bf 50}, 17953 (1994).
\bibitem{dft} P. Hohenberg and W. Kohn, Phys. Rev. {\bf 136}, B864 (1964) ; W. Kohn and L. J. Sham, ibid. {\bf 140}, A1133 (1965)
\bibitem{pbe} J. P. Perdew, K. Burke, and M. Ernzerhof, Phys. Rev. Lett. {\bf 77}, 3865 (1996) 
\bibitem{vesta} K. Momma and F. Izumi, J. Appl. Crystallogr., {\bf 41}, 653 (2008);, http://jp-minerals.org/vesta/en
\bibitem{moto-prb2017} M. Ezawa, Phys. Rev.  B {\bf 96}, 035425 (2017)
\end{thebibliography}
\end{document}